# Structural properties of TaAs Weyl semimetal thin films grown by molecular beam epitaxy on GaAs(001) substrates

*Janusz Sadowski[1,2,3]\*, Jarosław Z. Domagała[1], Wiktoria Zajkowska[1], Sławomir Kret[1], Bartłomiej Seredynski[2], Marta Gryglas-Borysiewicz[2], Zuzanna Ogorzalek[2], Rafał Bożek[2], and Wojciech Pacuski[2]*

[1] Institute of Physics Polish Academy of Sciences, Aleja Lotnikow 32/46, PL-02-668 Warsaw, Poland

[2] Institute of Experimental Physics, Faculty of Physics, University of Warsaw, Pasteura 5, PL-02-093 Warsaw, Poland

[3] Department of Physics and Electrical Engineering, Linnaeus University, SE-391 82 Kalmar, Sweden

ABSTRACT

Thin crystalline layers of TaAs Weyl semimetal are grown by molecular beam epitaxy on GaAs(001) substrates. The (001) planes of the tetragonal TaAs lattice are parallel to the GaAs(001) substrate, but the corresponding in-plane crystallographic directions of the substrate and the layer are rotated by 45°. In spite of a substantial lattice mismatch (about 19%) between GaAs(001) substrate and TaAs epilayer no misfit dislocations are observed at the GaAs(001)/TaAs(001) interface. Only stacking fault defects in TaAs are detected with transmission electron microscopy. Thorough X-ray diffraction measurements and analysis of the *in-situ* reflection high energy electron diffraction images indicates that TaAs layers are fully relaxed already at the initial deposition stage. Atomic force microscopy imaging reveals the columnar structure of the layers, with lateral (parallel to the layer surface) columns about 20 nm wide and 200 nm long. Both X-ray diffraction and transmission electron microscopy measurements indicate that the columns share the same orientation and crystalline structure.

INTRODUCTION

TaAs is the first experimentally recognized Weyl semimetal (WSM).[1,2] The possibility of condensed matter realization of formerly elusive Weyl fermions emerging as a zero mass solution of Dirac equation derived by Herman Weyl already in 1929,[3] generated recently a lot of interest in both theoretical[4,5,6,7] and experimental[1,2,8,9,10,11] solid state physics. WSM can be derived from slightly earlier known Dirac semimetals (DSM) by lowering necessary symmetry conditions fulfilled by the respective material.[8] In DSM both time reversal and inversion symmetry must be obeyed whereas in WSM one of those has to be broken.[8,12] WSM properties can occur in the material fulfilling above mentioned symmetry conditions if two bands without spin degeneracies cross in the close vicinity in the Brillouin zone; the crossing points are named Weyl nodes. Beside the linear dispersion relation of low energy excitations [Weyl fermions (WFM)] at the Weyl nodes the distinct (opposite) chiralities of WFM have pronounced consequences on magnetotransport phenomena in WSM. One of those is the occurrence of Adler-Bell-Jackiw (ABJ) anomaly which in WSM materials manifests as conductivity enhancement in parallel electric and magnetic fields, and appropriated orientation of both with respect to the WSM crystal lattice.[13,14] The TaAs crystal lattice fulfills time-reversal symmetry conditions but has no inversion symmetry. In the band structure of TaAs 12 pairs of Weyl nodes have been identified in the distinct points of the Brillouin zone;[4] but So far TaAs has been obtained only in the form of bulk crystals. In this report we show that TaAs can also be grown by molecular beam epitaxy (MBE) on commonly used GaAs(001) substrates. This enables integration of this WSM in heterostructures with ferromagnetic and/or antiferromagnetic materials such as (Ga,Mn)As,[15] MnAs,[16] CuMnAs,[17] also possible to grow on GaAs(001) and get use of magnetic proximity effects instead of the external magnetic field to attain distinct magnetotransport phenomena characteristic for TaAs[18] and other WSM materials.[19]

RESULTS AND DISCUSSION

TaAs layers are grown by MBE on GaAs(001) substrates with predeposited GaAs buffer layers grown in the same MBE system (see methods for details). The growth was monitored *in-situ* by reflection high energy electron diffraction (RHEED). For all grown TaAs layers 2-dimensional RHEED patterns were observed from the deposition of the very first monolayer and persisted during further growth. Analysis of the spacing of TaAs electron diffraction streaks (see Fig. 1) indicates a full relaxation of TaAs epilayer already at the initial growth stage. TaAs crystallizes in a body-centred-tetragonal structure with the lattice parameters: a = b = 3.4368 Å; c = 11.6442 Å.[20] In spite of a substantial lattice mismatch (about 19%) between the in-plane directions of GaAs(001) substrate and corresponding directions of TaAs epilayers there are no traces of spotty RHEED features pointing on islanding, i.e., 3-dimensional growth mode, at any stage of the TaAs MBE growth. The comparison of the spacing of RHEED streaks originating from the GaAs substrate with those from TaAs layer yields the in-plane lattice parameter value of the latter equal to 3.5 ± 0.1 Å, which coincides well with the $a$ and $b$ lattice parameters of TaAs bulk crystal. This in-plane TaAs lattice parameter value deduced from RHEED patterns along [$\bar{1}$10] and [110] azimuths of the GaAs(001) substrate indicates that TaAs grows with (001) planes parallel those of the GaAs substrate but the corresponding azimuths of the (001) planes of the substrate and the layer are rotated by 45 degrees.

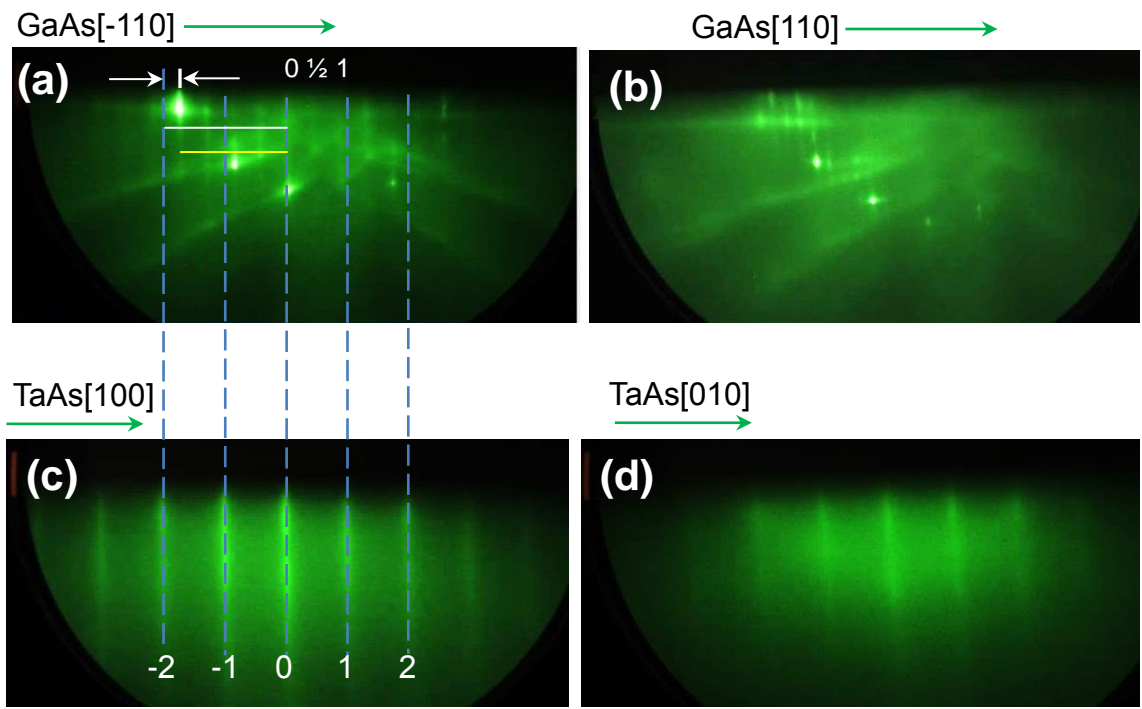

**Figure 1.** RHEED images of GaAs(001) substrate for two orientations rotated by 90 degrees – (a), (b); and for TaAs epilayer (c), (d) after 26 min of the growth corresponding to about 1.5 ML of TaAs. Green arrows mark the crystallographic directions; blue dashed lines are plotted for comparison of TaAs diffraction streaks spacing with respect to those of the GaAs(001) substrate.

It is clearly visible that the intensities of TaAs electron diffraction streaks are much higher along the [$\bar{1}$10] azimuth of the GaAs(001) substrate than those along 90 degree rotated [110] one. This can point on some disorder in the TaAs layer in this direction. Indeed the surface morphology of TaAs layer visualized by atomic force microscopy (AFM) displayed in Fig. 2 reveals the occurrence of lateral columns parallel to the [$\bar{1}$10] azimuth of the substrate. Such surface morphology of TaAs layer grown on GaAs(001) was observed for all the samples investigated by us (about 10 samples), irrespectively of the growth conditions such as substrate temperature and As/Ta flux ratio.

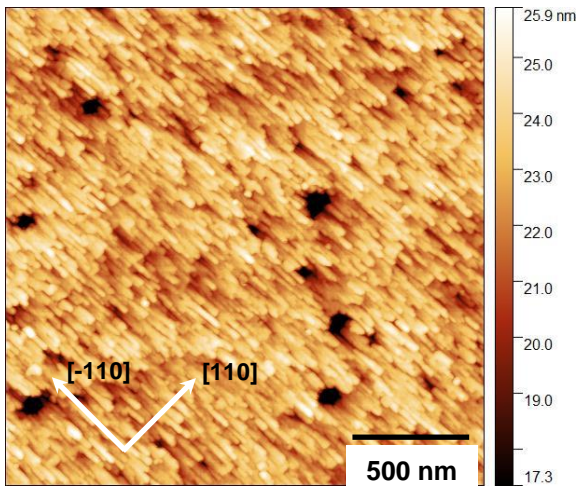

**Figure 2.** AFM surface morphology of a 9 nm thick TaAs layer grown on GaAs(001). RMS roughness of the TaAs area shown in the picture is equal to 1.5 nm.

Structural properties of the TaAs epilayers are investigated by X-ray diffraction (XRD) enabling determination of the crystallographic structure of the layer, its quality, roughness (surface and interface roughness); thickness, and lattice parameters. For XRD investigations we have chosen 18 nm thick TaAs layer, so two times thicker than that shown in Fig.2.

Figure 3 displays the X-ray reflectivity curve (intensity beam collection versus coupled sample and detector move with the same angular speed, for very small incident angles, from 0 to 3-6°); compared with the respective simulation curve obtained with the Panalytical X'Pert Reflectivity 1.3a software. Clearly visible oscillations of the reflectivity curve decreasing from the critical angle ($\omega_{crit} \sim 0.36°$) and decaying in the vicinity of $2°$, can be well simulated using the following parameters: $d_{layer}$ = 18.1 nm, electron density equal to 11.3 g / cm$^3$, surface roughness of 1.9 nm and the interface roughness of 0.95 nm.

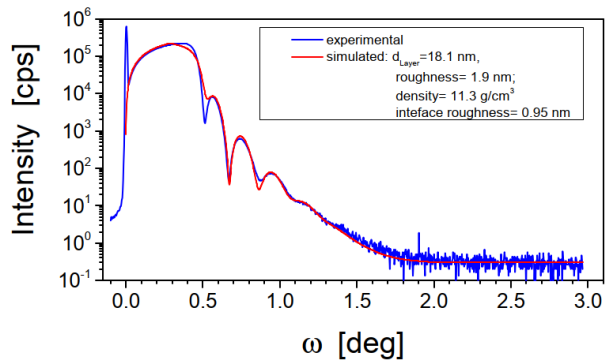

**Figure 3.** X-ray reflectivity curves from 18.1 nm thick TaAs/GaAs(001) epilayer, experimental (blue) and simulated (red).

Wide range of the 2θ/θ scan in the sample adjustment to the (001) GaAs plane was chosen to verify if the TaAs layer has only one crystalline phase and has an epitaxial relation to the substrate. The result is shown in Fig. 4a (red labels mark the peaks of the layer, blue labels – those of GaAs). The positions of the layer peaks perfectly match the successive reflections (four, eight and twelve) from the (001) TaAs plane with an interplanar distance of 11.705 Å. There are no other peaks indicative of other phases. The (001) plane of the layer is parallel to the (001) one of GaAs.

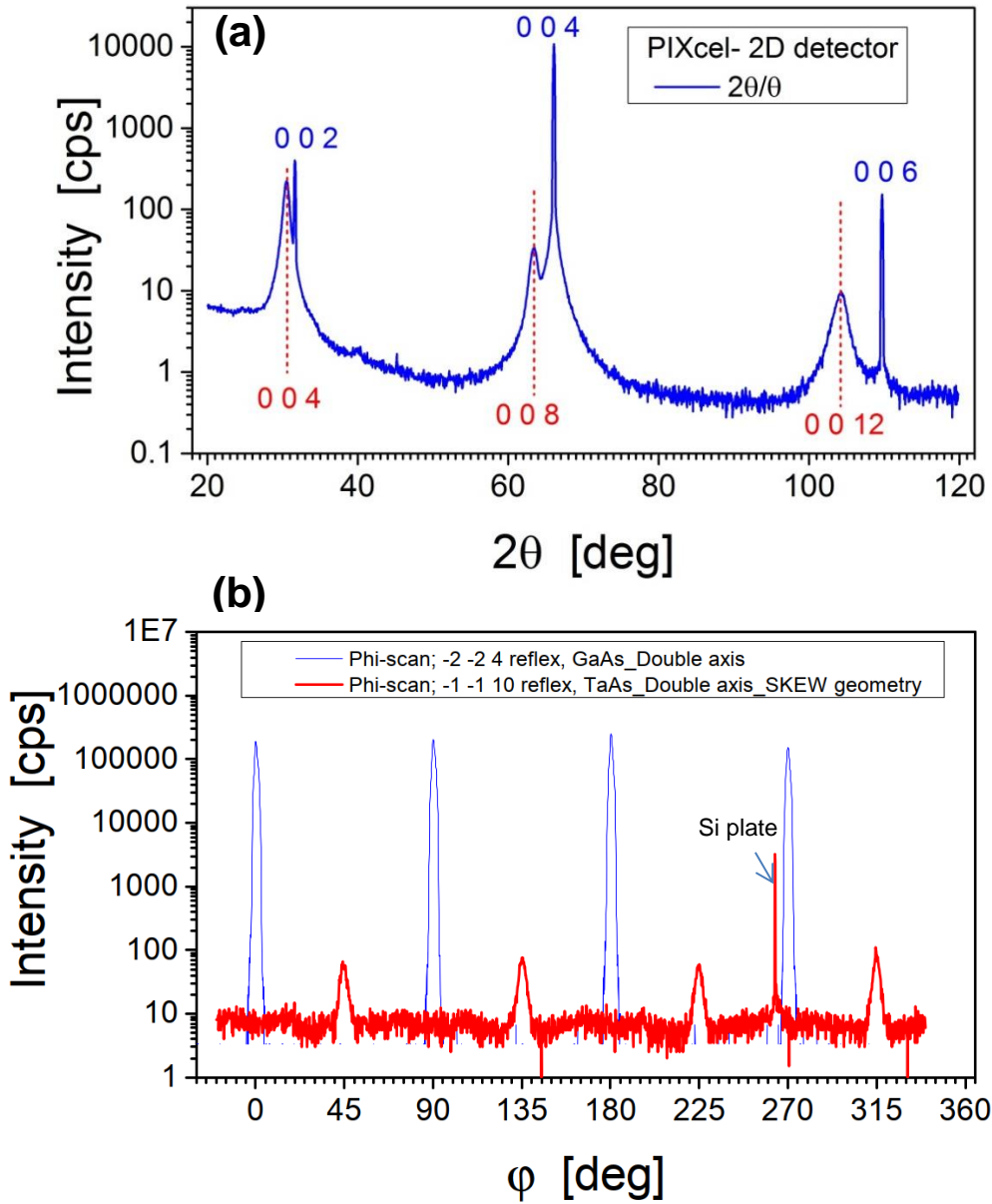

**Figure 4.** (a) - 2θ/θ curve with marked identifiers of reflections for the GaAs(001) substrate (upper blue indices) and 18.1 nm thick TaAs(001) layer (lower red indices); (b) - X-ray φ scans,

blue curve: $\bar{2}\bar{2}4$ GaAs reflection, red solid curve $\bar{1}\bar{1}10$ TaAs reflection (skew geometry); red peak around φ ~ 264⁰ is a signal from the Si plate of the diffractometer sample holder.

The angular positions of the layer diffraction peaks is consistent with those described in Ref. [20] - the Powder Diffraction File (PDF) 33-1388 for TaAs; i.e. tetragonal crystal system, space group I41md (nr 109) and lattice parameters: a=3.4368 Å, c=11.6442 Å and density 12.35 g/cm$^3$. Using the PDF 33 -1388 data, it is easy to calculate the angular positions of the detector and sample settings for measuring asymmetric $\bar{1}\bar{1}10$ reflection. Despite the very small layer thickness (18 nm), we have managed to record reciprocal space map (RSM) for this reflection

The directional relationships at the layer-substrate interface can be illustrated by measuring the intensity of reflected X-ray beam for the measurable asymmetric reflections of TaAs - here: $\bar{1}\bar{1}10$ and GaAs ($\bar{2}\bar{2}4$) versus rotation around the axis perpendicular to the sample surface i.e. [001] GaAs direction (see Fig.4b). Due to the very small TaAs layer thickness, the $\bar{1}\bar{1}10$ reflex is measured in the so-called skew geometry. In this geometry, the angle of incidence of the beam is equal to the Bragg angle and the normal to the surface does not lie in the diffraction plane (defined by the wave vectors of the incident beam and the reflected beam). Maxima corresponding to the GaAs substrate and TaAs layer are shifted by 45° which is a consequence of 45° rotation of the corresponding <110> directions of the layer and the substrate.

The lattice unit parameters of the TaAs layer (c=11.705±0.003 Å, a=3.447±0.005 Å) are determined from the RSMs of two 0 0 12 and $\bar{1}\bar{1}10$ reflections measured with 2D-Pixcel detector in the scanning mode (255 active stripes); see Fig 5 (a,b), respectively. Maps for other symmetric reflections - (004, 008), are also recorded. The layer spots displayed in Figure 5 are weak in intensity and extended in the directions perpendicular and parallel to the diffraction vector q (shown as a red, dotted line). It means that we observe a kind of mosaicity (perpendicular broadening) and relatively large spread of d-interplanar space parameters (radial broadening) in the TaAs layer.

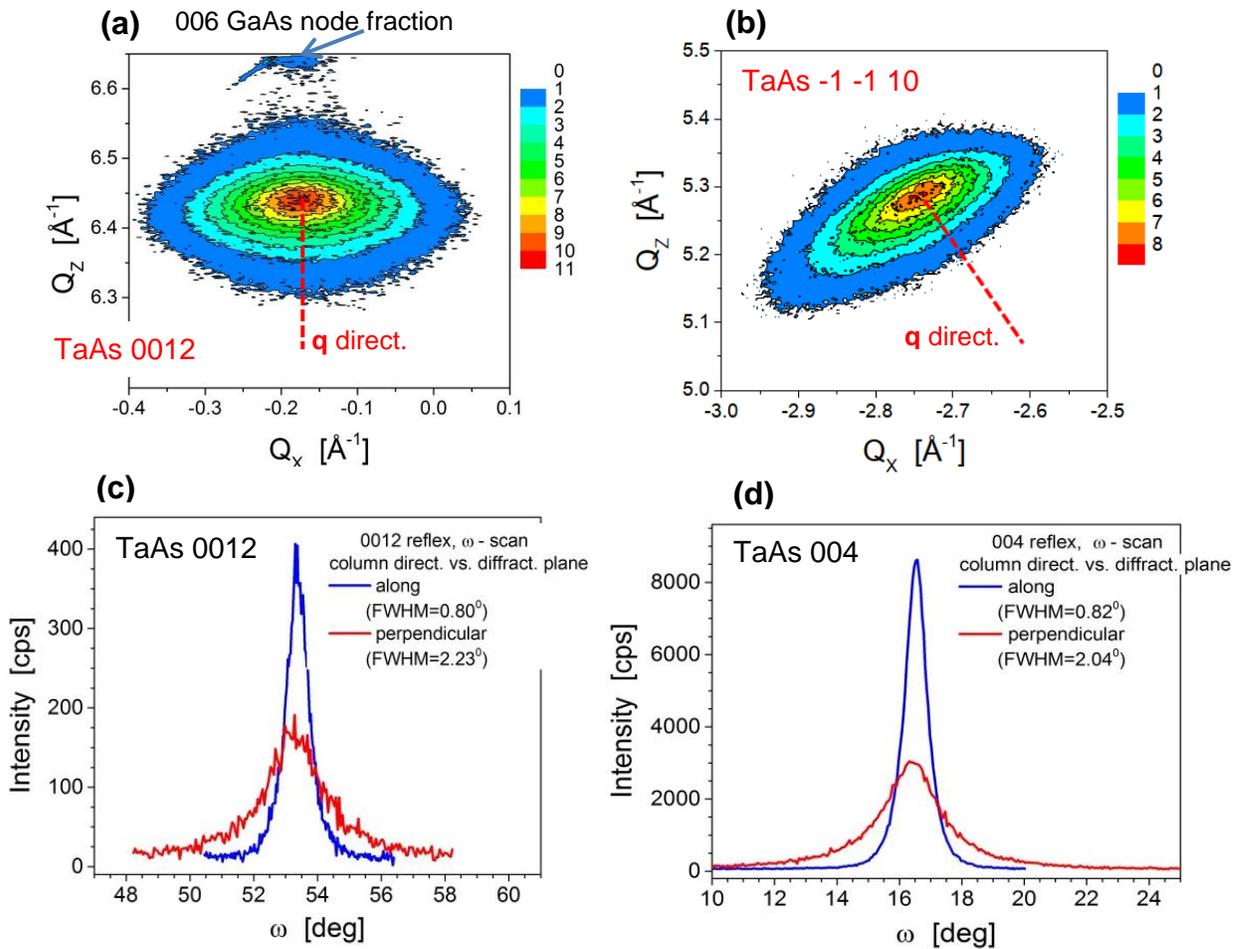

**Figure 5.** (a) - RSM of symmetrical TaAs 0 0 12 reflection recorded for the 18.1 nm thick TaAs layer, at the top – a fraction node of GaAs 006 is visible; (b) - RSM of asymmetrical TaAs $\bar{1}\bar{1}10$ reflection, red dotted lines show direction of the diffraction vector; (c, d) - X-ray rocking curves of 0 0 4, and 0 0 12 reflections from TaAs layer, measured in two directions perpendicular to the position of the layer columns, FWHM sizes of the curves are given.

Interestingly both *a* and *c* TaAs lattice parameters measured by us are slightly higher (0.3% and 0.52%, respectively) than those of the TaAs bulk crystals. Similar effect has been observed by A. Bedoya-Pinto et. al.[21] for NbP (with tetragonal lattice and Weyl semimetal properties, both similar to those of TaAs) grown on MgO(001) substrates. Moreover the 45 degree in-plane rotation of the crystal lattice of the films with respect to the substrate has been noticed for NbP/MgO(001) akin to the TaAs/GaAs(001) case.

Typically, the quality of a layer is assessed by full widths and half maxima (FWHM) of the rocking curves of symmetric reflections. Analysis of the 2θ/θ scan shown in Figure 4a, reveals that the 004 and 0012 reflections peaks are not much disturbed by (well separated from) the substrate peaks, so the rocking-curve (RC) shapes for these reflections depend only on the quality of the layer. In Figure 5 (c, d) we show the experimental RC of the 18 nm thick TaAs layer for the 004 and 0012 reflections. The FWHM of the corresponding RC curve depends on the position of the lateral columns clearly visible at the AFM image displayed in Fig. 2 (18.1 nm thick TaAs has the same surface morphology as the 9 nm film shown in Fig. 2). When the columns are parallel to the diffraction plane FWHM is relatively small (~ 0.8°), while when they are perpendicular - it is nearly 3 times larger (~2.1°). Variations of RC FWHM, (in our case, increase from 0.8° to 2.2° for symmetric reflections in the two orthogonal sample position twisted around the surface normal) can occur in heteroepitaxial systems and were associated previously with the misfit dislocation anisotropy in characteristic directions.[22,23] Obviously the signal from a very thin layer always introduces additional broadening of the reflection curves, especially 2θ/θ one and RSM in the radial direction with respect to q [see Fig. 4a and Fig.5 (a,b)]. In our case this effect on the FWHM of 2θ/θ peak (see Fig.4) for 0012 peak amounts to about 0.7° (006 reflection simulations for an ideal 18.1 nm thick InAs layer grown on GaAs(001); simulation program PANalytical Epitaxy 4.3a) which is about 3 times smaller than FWHM (2.02°) of the experimental TaAs peak. It means that we observe some imperfections varying the (001) interplanar distances. Thus, taking into account common assumption that the square of the width of the diffraction curve is related to the sum of the squares of the natural widths of the material and the mosaic, it is easy to calculate the effect of the latter. One can see that more than half of the radial expansion of RSMs (1.6°) is related to defects changing the interplanar distance, e.g. stacking faults. Indeed the stacking fault defects have been observed in TaAs bulk crystals.[24] The ellipsoidal shapes of the RSM spots (see Figures 5a and 5b) mean that the corresponding layers are extended not only in the radial direction of the diffraction vector **q** but also in the direction perpendicular to it. This phenomenon may be due to the columnar structure of the layer (perfect regions (columns) slightly misoriented with respect to each other); clearly visible in the AFM surface morphology of TaAs layer shown in Figure 2. The orientation of lateral TaAs nanocolumns along the [$\bar{1}$10] direction of GaAs(001) substrate surface is apparently associated with the anisotropy of (2x4) reconstructed GaAs(001). Even prolonged annealing of (2x4) GaAs(100) in As flux can lead to emergence of nanoislands elongated in this direction, as shown by La Bella et. al.[25] Moreover deposition of metals (e.g. Fe) on GaAs(001) can result in the stripe like morphology with elongated surface features parallel to GaAs [$\bar{1}$10].[26] Interestingly in the reported MBE growth of TaP, with the crystalline structure identical to that of TaAs, on MgO(001) substrate without the surface anisotropy similar to GaAs(001), the block-like structure in TaP films was observed but with no preferential elongation along any MgO(001) surface azimuth.[21]

The local crystalline structure of TaAs/GaAs interface and TaAs layer has been investigated by high resolution electron microscopy (TEM) in both conventional imaging and scanning mode (STEM), using cross-sectional specimens prepared with the focused ion beam technique. Two cross-sections have been fabricated, rotated by 45 deg to one another. TEM examination confirms that the crystall lattice of TaAs layer belong to the I41md space group. The azimuthal orientations of the TaAs layer with respect to the GaAs(001) substrate are in accordance with the XRD results and are identified in TEM images as follows: TaAs[1$\bar{1}$0] || GaAs[0$\bar{1}$0] – see Fig.6 (a,b); TaAs[0$\bar{1}$0] || GaAs[1$\bar{1}$0] – see Fig.6 (c,d). The epitaxial relation of TaAs layer to the GaAs(001) substrate is evidenced, as shown in Figs 6(a,b) and 6(c,e), at low and high magnification, respectively.

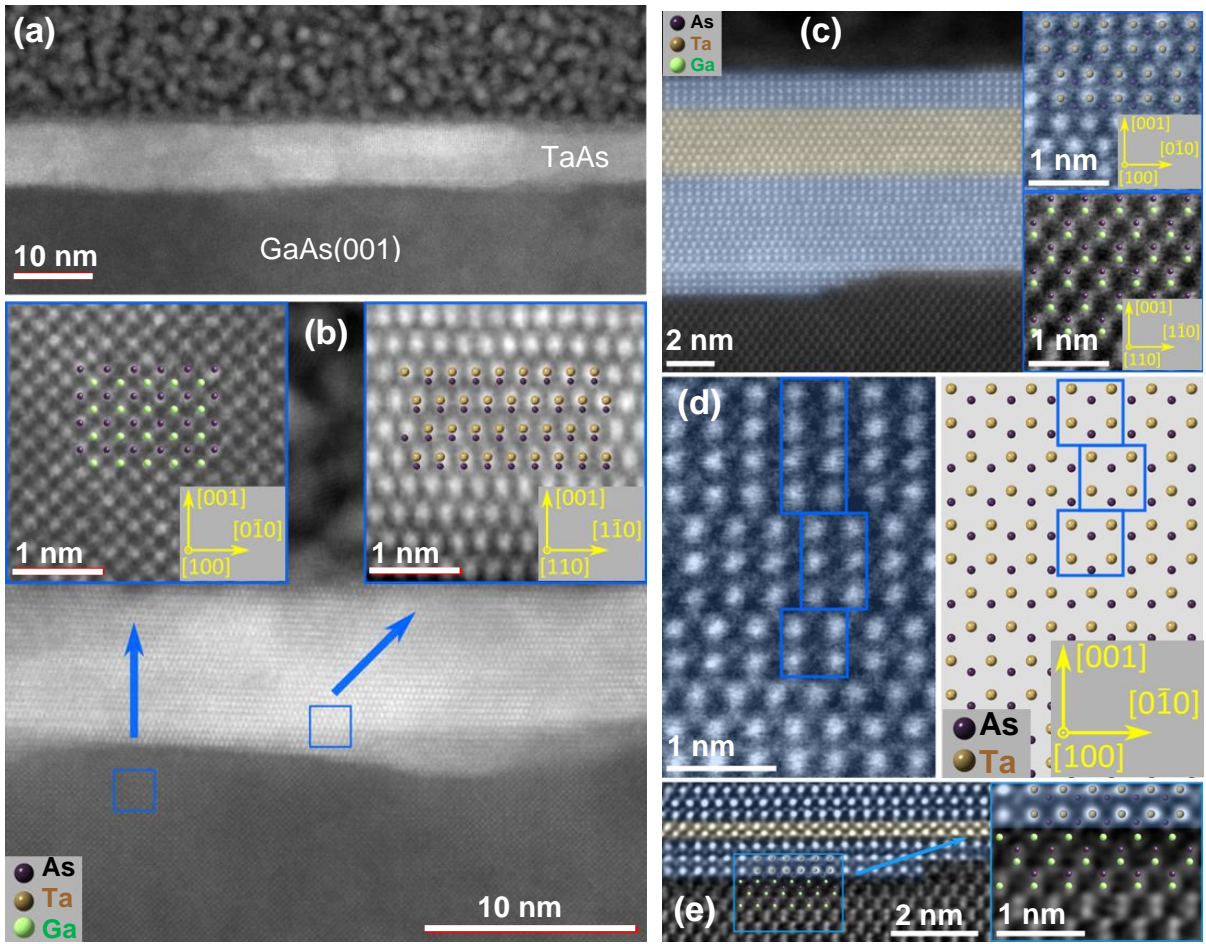

**Figure 6.** STEM image of TaAs[1$\bar{1}$0] || GaAs[0$\bar{1}$0] interface (9 nm thick layer) (a) - in low magnification, (b) - with zoomed GaAs and TaAs areas; (c-e) image of TaAs[0$\bar{1}$0] || GaAs[1$\bar{1}$0]

interface; zoomed areas of each layer are shown in the insets to (c); (d) - stacking faults in the TaAs layer; (e) proposed reconstruction of the GaAs-TaAs interface.

Misfit dislocations are not visible neither along the [1$\bar{1}$0] direction nor along the [0$\bar{1}$0] one of TaAs. Additionally, the GaAs and TaAs structures (generated with the Crystal Maker software) are placed on the STEM images in the corresponding crystal orientations. The images shown in Fig. 6 (a,b) are taken in the mass contrast mode (camera length equal to 73 mm) and in Fig. 6 (c,e) in the mixed mass-diffraction contrast (camera length equal to 185 mm). Due to that and because of the fact that As atoms are much lighter than Ta ones, As is hardly visible in the TaAs layer. The atomic planes order shifts in TaAs can be seen in Fig. 6d. Not every second, but every 3-rd or 4-th layer is laterally shifted along the [0$\bar{1}$0] direction with respect to the underlying one. Moreover, in Fig. 6c the orientation changes of TaAs crystal are evidenced. In the area close to the interface, the TaAs layer is visible in the [100] zone axis parallel to the [110] one of the GaAs substrate. In the area of the layer located ca. 4 nm above the GaAs substrate, one can observe that TaAs(001) planes are rotated by 45° so they are seen in the [0$\bar{1}$0] direction. These planes (marked in yellow in Fig. 6c) clearly differ from the ones with prevailing orientation marked in blue. Additionally, near the layer surface, the orientation of TaAs changes again by 45° rotation. In Fig. 6c one can observe a step in the GaAs//TaAs interface what is a consequence of the substrate miscut. Figure 6e shows the GaAs//TaAs interface with its reconstruction model (superimposed position of atoms). It is worth to underline that despite the local azimuthal orientation rotations revealed in TEM images the entire TaAs layer grew along [001] direction with prevailing lateral orientation of TaAs(001) lattice rotated (in the layer plane) by 45 degree with respect to the GaAs(001) substrate.

Due to the nanostripe morphology of TaAs layers, we have investigated a possible impact of this structure on the electric properties of the layers. Two sets of macroscopic hallbars were prepared for electric measurements (with current path 600 μm in length and 200 μm in width), along directions parallel and perpendicular to the nanostripes, corresponding to [$\bar{1}$10] and [110] directions. The anisotropy of resistivity is observed, with 3.5 times lower resistivity for the current path along the nanostripes ($\rho_{\parallel}$), in comparison to the perpendicular one ($\rho_{\perp}$). The room temperature resistivities are equal to: $\rho_{\parallel} \sim 20$ μΩ∗cm and $\rho_{\perp} \sim 72$ μΩ∗cm, respectively. These values are in a good agreement with the data for TaAs bulk samples obtained by chemical vapor method.[27]

CONCLUSIONS

In conclusion we have demonstrated a successful MBE growth of TaAs Weyl semimetal thin crystalline layers on commonly used GaAs(001) substrate. TaAs films share the (001) orientation with the GaAs substrate but they are in-plane rotated by 45 degree; i.e. the <010> directions of a TaAs layer are parallel to the <011> directions of the GaAs substrate. In spite of a large lattice mismatch between both crystalline lattices (about 19% in such orientation) TaAs grows in a 2D layer-by-layer mode from the very first monolayer and no misfit dislocations in the TaAs/GaAs interface are identified in the cross-sectional TEM images. However, some mosaicity of the TaAs layer is revealed by the X-ray diffraction measurements and is manifested by a stripe-like TaAs surface morphology with about 20 nm thick and 200 nm long TaAs stripes all sharing the same crystallographic orientation parallel to the in-plane [$\bar{1}$10] direction of the GaAs(001) substrate. At higher TaAs layer thicknesses (above about 5 nm) the columns merge together and form a continuous layer. Our results demonstrate the possibility of integrating monocrystalline layers of TaAs Weyl semimetal investigated so far in the form of bulk crystals, with other materials such as ferromagnets, antiferromagnets or superconductors in all epitaxial heterostructures, which opens a way for investigations of a plethora of new phenomena involving proximity effects with topological Weyl semimetals.

METHODS

**Growth of the samples**. TaAs layers are grown in a III-V MBE system (SVTA) equipped with the valved-cracker source for As and e-beam evaporator for Ta. The ¼ of the 3 inch GaAs(001) wafers have been used as substrates. After standard procedures of thermal native oxide desorption and about 150 nm thick GaAs buffer layer growth, the TaAs layer was grown at the same (high) temperature as used for GaAs buffer (about 600 °C), and with the same $As_2$ flux intensity. Other growth conditions - lower substrate temperature and/or $As_2$ flux were observed to yield poorer quality of TaAs, as evaluated in-situ by RHEED. The growth time of a 9 nm thick TaAs film was 1h, the 18 nm thick one was grown for 12 hours with 1 order of magnitude lower Ta flux. No essential differences in the surface morphologies and crystalline quality of both films were observed.

**X-Ray Diffraction**. Two high resolution diffractometers equipped with a Cu X-ray tube, Panalytical Empyrean one and Philips X'Pert MRD were used in the experiment. Lattice parameters were calculated on the basis of reciprocal space map (RSM) measurements taken using the Empyrean with horizontal sample stage, high intensity beam flux (hybride two-bounce Ge (220) monochromator) and fast 2D detector PIXcel. High X-ray beam intensity used for the measurements and the strip structure of the detector made it possible to efficiently scan a large range of the reciprocal lattice space in order to identify relatively weak asymmetric reflections from the thin TaAs layer. X'Pert MRD diffractometer operating in the high-resolution mode (4-reflection monochromator Ge (220) with X-ray mirror) and with the Eulerian sample holder, allowed to measure pole figures (series of Phi scans at regularly spaced Psi positions) in order to establish dependence of layer and substrate directions in the interface plane. For both diffractometers, the monochromators formed a beam with a wavelength of 1.5406 Å (CuK$_{\alpha1}$) from the X-ray spectrum of the tube.

**Transmission electron microscopy**. The investigation of the samples was performed using Titan Cubed 80-300 transmission electron microscope (TEM) operating at 300 kV and equipped with an image corrector. The scanning transmission electron microscopy high angle annular dark field (STEM−HAADF) images were acquired with 73 and 185 mm camera length. The upper scattering angles for electrons were cut-off due to the image corrector use.

In order to prepare two specimens rotated by 45 degrees to each other a Helios Nanolab 600 microscope with Focused Ion Beam (FIB) system was used. Electron transparent lamellae were cut with gallium ions and transferred by an Omniprobe nanomanipulator to a copper grid compatible with a TEM holder.

AUTHOR INFORMATION

**Corresponding Author**

* E-mail: janusz.sadowski@lnu.se

**Author Contributions**

The manuscript was written through contributions of all authors. All authors have given approval to the final version of the manuscript.

**Notes**

Very recently we became aware of the reported MBE growth of polycrystalline TaAs films on GaAs(001) substrates, also interfaced with NiFe ferromagnetic layers.

Yanez, Wilson; Ou, Yongxi; Xiao, Run; Ghosh, Supriya; Dwivedi, Jyotirmay; Steinebronn, Emma; Richardella, Anthony; Mkhoyan, K. Andre; Samarth, Nitin. Giant spin torque efficiency in naturally oxidized polycrystalline TaAs thin films. **arXiv:2202.10656**

ACKNOWLEDGMENTS

This work has been supported by the National Science Centre (Poland), through the projects No. 2017/27/B/ST5/02284, 2021/41/ST3/04183 and by the Swedish Research Council, through the project No. 2017-04404.

**For Table of Contents Use Only**

Structural properties of TaAs Weyl semimetal thin films grown by molecular beam epitaxy on GaAs(001) substrates.

Janusz Sadowski, Jarosław Z. Domagała, Wiktoria Zajkowska, Sławomir Kret, Bartłomiej Seredynski, Marta Gryglas-Borysiewicz, Zuzanna Ogorzalek, Rafał Bożek, and Wojciech Pacuski

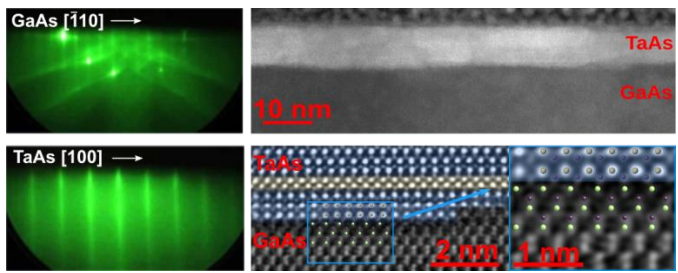

Structural properties of TaAs Weyl semimetal thin films grown by molecular beam epitaxy on GaAs(001) substrates are studied by X-ray diffraction (XRD) and electron transmission microscopy (TEM). In-situ reflection high energy electron diffraction as well as XRD and TEM measurements point on the monocrystalline structure of the films, however the stripe-like TaAs surface morphology is revealed by atomic force microscopy.